\documentclass[aps,twocolumn,letter,showpacs,prepr intnumbers,amsmath,amssymb,superscriptaddress,bibnotes]{revtex4-1}

\usepackage{graphicx}
\usepackage{dcolumn}
\usepackage{bm}
\usepackage{textcomp}
\usepackage{amsmath}
\usepackage{amssymb}
\usepackage{color}
\usepackage{soul}
\usepackage{natbib}
\usepackage{dsfont}

\begin{document}

\title{Strongly Temperature-Dependent Spin-Orbit Torques in Heavy Fermion YbAl$_3$}
\author{Neal D. Reynolds} 
\email{ndr37@cornell.edu}
\thanks{equal contribution}
\affiliation{Department of Physics, Cornell University, Ithaca, NY 14853, USA}
\author{Shouvik Chatterjee}
\email{sc2246@cornell.edu}
\thanks{equal contribution}
\affiliation{Department of Physics, Cornell University, Ithaca, NY 14853, USA}
\author{Gregory M. Stiehl}
\affiliation{Department of Physics, Cornell University, Ithaca, NY 14853, USA}
\author{Joseph A. Mittelstaedt}
\affiliation{Department of Physics, Cornell University, Ithaca, NY 14853, USA}
\author{Saba Karimeddiny}
\affiliation{Department of Physics, Cornell University, Ithaca, NY 14853, USA}
\author{Alexander J. Buser}
\affiliation{Department of Physics, Cornell University, Ithaca, NY 14853, USA}
\author{Darrell G. Schlom}
\affiliation{Department of Materials Science and Engineering, Cornell University, Ithaca, NY 14853, USA}
\affiliation{Kavli Institute at Cornell, Cornell University, Ithaca, NY 14853, USA}
\affiliation{Leibniz-Institut f\"{u}r Kristallz\"{u}chtung, Max-Born-Str. 2, 12489 Berlin, Germany}
\author{Kyle M. Shen}
\affiliation{Department of Physics, Cornell University, Ithaca, NY 14853, USA}
\affiliation{Kavli Institute at Cornell, Cornell University, Ithaca, NY 14853, USA}
\author{Daniel C. Ralph}
\affiliation{Department of Physics, Cornell University, Ithaca, NY 14853, USA}
\affiliation{Kavli Institute at Cornell, Cornell University, Ithaca, NY 14853, USA}

\begin{abstract}
The use of current-generated spin-orbit torques to drive magnetization dynamics is under investigation to enable non-volatile, low-power magnetic memory. Previous research has focused on spin-orbit torques generated by heavy metals, interfaces with strong Rashba interactions, and topological insulators, which can all be well-described using models with noninteracting-electron bandstructures.  Here, we show that electronic  interactions  within a strongly correlated Kondo lattice system YbAl$_{3}$, can provide a large enhancement in spin-orbit torque. The spin-torque conductivity increases by approximately a factor of three from room temperature to the coherence temperature of YbAl$_{3}$ (T* $\sim$ 37 K), which mimics the temperature dependence of the density of states at the Fermi level. This indicates that the renormalization of electron bands associated with the many-body Kondo resonance provides a large enhancement of spin-orbit torque. Our observation suggests new opportunities in spin-orbit torque manipulation by utilizing quantum many-body states.
\end{abstract}

\maketitle


Previous research on materials that can be used to generate spin-orbit torques \cite{Sinova2015}  has focused primarily on materials containing $d$-electrons, including heavy metals \cite{ando2008electric,Liu2011,miron2011perpendicular,Liu2012,Pai2012,Emori2013,Ryu2013}, interfaces with strong Rashba interactions \cite{Jungfleisch2016, yue2018spin}, and topological insulators \cite{Mellnik2014,fan2014magnetization,khang2018conductive,Dc2018}, with only a few exceptions \cite{Tanaka2009,Tanaka2010,Singh2015, Ueda2016, Reynolds2017, li2018thin}. In $d$-electron materials, inter-atom coupling produces electronic bands with relatively large bandwidths, such that effective independent-electron models provide a reasonable description of most  electronic properties.  In $f$-electron materials, by comparison, the electron bandwidths are  narrower and electronic interactions can be much more impactful, increasing the potential importance of electron correlations.  For the pure $f$-electron elements measured to date,  hybridization between the electron states at the Fermi level and the $f$-electron levels happens to be negligible (the dominant contribution to the electronic density of states near E$_{F}$ is from the $d$ orbitals), and hence the effects of electron correlations on the spin-orbit torques from these materials are expected to be only modest \cite{Reynolds2017}. Previous studies of pure $f$-electron elements have measured \cite{Singh2015, Ueda2016, Reynolds2017} a maximum spin-torque conductivity at room temperature of $2 \times 10^4$ $\frac{\hbar}{2e}$ $\Omega^{-1}$ m$^{-1}$  for holmium (Ho) \cite{Reynolds2017}.

To probe whether correlated-electron effects in $f$-electron materials can enhance spin-orbit torques, we investigated the Kondo lattice system YbAl$_{3}$ (Fig.~1(a)).  YbAl$_{3}$ is a well-studied material \cite{cornelius2002two, ebihara2003dependence} with a relatively high Kondo temperature of $\approx$ 670 K, the energy scale for the Kondo interaction between the localized Yb 4$f$ moments and the delocalized conduction electrons that leads to the formation of a Kondo screened many-body state of predominantly Yb 4$f$ character. As the Kondo interaction strength increases with decreasing temperature, it results in the emergence of a narrow Yb 4$f$-derived  band close to the Fermi level that is prominent in angle-resolved photoemission spectroscopy (ARPES) measurements \cite{Chatterjee2017}. Therefore, unlike previous measurements on $f$-electron systems, YbAl$_{3}$ offers a unique opportunity to investigate the role of renormalized $f$ states on the generation of large spin-orbit torque by varying the sample temperature. Figures 1(d) and 1(e) show the evolution of ARPES spectra between 255 K and 21 K for 20 nm thick YbAl$_{3}$ thin films  grown by molecular-beam epitaxy (MBE). Near the Fermi energy, the integrated spectral weight averaged over a region of momentum space away from the electron pocket centered at $k_{\parallel}$=0  [\textit{i.e.} the momentum region where only the Yb-4$f$-derived Kondo band is present (see section 5.3 in the Methods section)] grows approximately logarithmically  with decreasing temperature from 250 K down to the coherence temperature ($T^* \approx 37$ K), and saturates at lower temperatures where YbAl$_{3}$ exhibits Fermi liquid behavior (Figs.~1(c),1(h),1(j)). This observation is in accord with expectations from a phenomenological two-fluid model, as has also been seen in other Kondo lattice systems \cite{choi2012temperature,yang2008universal,yang2012emergent}. 

\begin{figure*}[!tbp]
\includegraphics[width=1\textwidth]{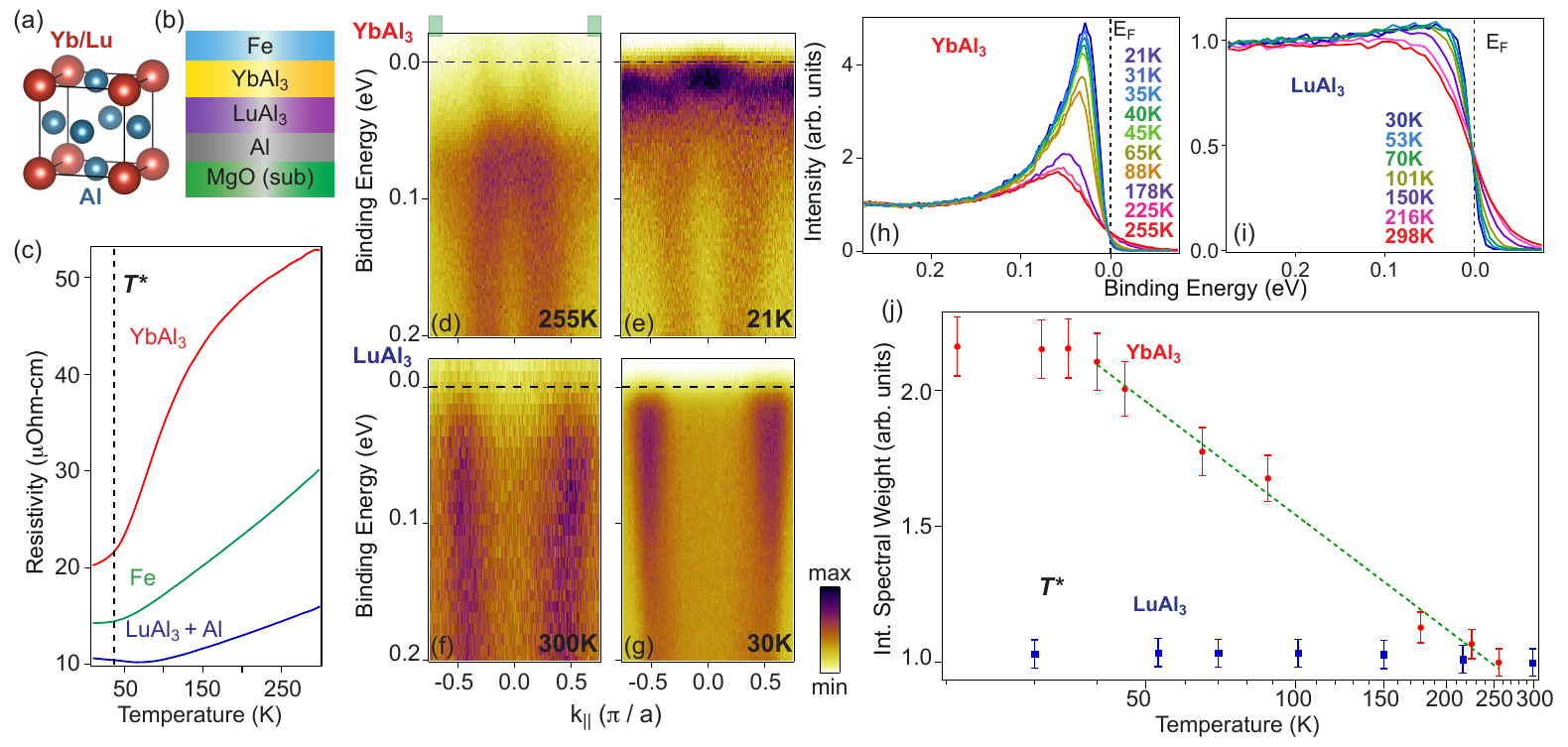}
\centering
   \caption{
   \textbf{Transport and spectroscopy of YbAl$_{3}$ and LuAl$_{3}$ thin films.} (a) Crystal structure of YbAl$_{3}$ and LuAl$_{3}$. (b) Sample structure used for the ST-FMR and MOKE measurements. For the ARPES measurements, the Fe layer was absent. (c) Temperature-dependent resistivities of YbAl$_{3}$, LuAl$_3$ + Al, and Fe. The dashed line indicates the coherence temperature of YbAl$_3$ ($T^* = 37$ K) below which its resistivity exhibits a Fermi-liquid behavior, $\rho \propto T^{2}$. (d-g) Measured \textit{E-k} spectral maps along (0,0) - (0,$\pi$) near the bulk $\Gamma$ point for YbAl$_{3}$ at (d) 255 K and (e) 21 K,  and for LuAl$_{3}$ at (f) 300 K and at (g) 30 K. (h,i) Energy distribution curves (EDCs) at different sample temperatures for (\textbf{H}) YbAl$_3$ integrated over a momentum region   $\pm$(0.66 - 0.75 $\pi/a$) indicated by the green rectangles at the top of (d) and for (i) LuAl$_{3}$ over a momentum region -0.75$\pi/a$ and 0.75 $\pi/a$. (j) Spectral weight integrated over $-$0.05-0.2 eV binding energy for (red circles) YbAl$_{3}$ and (blue squares) LuAl$_{3}$, after subtracting an inelastic background.}
 \label{fig1}
\end{figure*}

Here, we report measurements of the strength of spin-orbit torque produced by YbAl$_{3}$ as a function of temperature to study whether the Kondo-induced bandstructure renormalization, as observed via ARPES, can enhance the spin-torque efficiency. In order to distinguish the influence of 4\textit{f} states involved in the YbAl$_{3}$ Kondo resonance from the contributions of other $d$, $p$, and $s$ derived bands, we also examined control samples of LuAl$_{3}$, which has a similar crystal structure and lattice constants as YbAl$_{3}$, but in which the Kondo interaction is absent due to the fully-filled 4$f$ shell of the Lu$^{3+}$ ions. Therefore, unlike in YbAl$_{3}$, only wide bands predominantly of Al $p$ orbital character  cross the Fermi level in LuAl$_{3}$ (Figs.~1(f),1(g)), with negligible temperature dependence for the integrated spectral weight (Figs.~1(i),1(j)).



To measure the spin-orbit torque generation in these compounds, we  fabricated heterostructures in which epitaxial layers of ferromagnetic Fe were grown on single crystal epitaxial layers of YbAl$_{3}$ or LuAl$_3$. The growth was performed by molecular-beam epitaxy on MgO substrates as described in section I in the supplementary information and in ref.~\cite{Chatterjee2016}. To obtain high quality heterostructures, atomic layers of YbAl$_{3}$ were first deposited on buffer layers containing 1.8 nm thick Al and 5 nm thick LuAl$_3$ atomic layers. Samples for ARPES were measured \textit{in situ} after deposition of 20 nm YbAl$_3$ or 30 nm LuAl$_3$. The samples analyzed for spin-torque measurements contained 1.8 nm Al, 5 nm LuAl$_{3}$, (6.7, 15, 18) nm  YbAl$_{3}$, and 6 nm Fe, while the control samples consisted of 1.8 nm Al, 15 nm LuAl$_{3}$, and 6 nm Fe. The spin-torque samples were capped with 3 nm of Al, which oxidizes to form a protective layer of AlO$_x$ upon exposure to atmosphere. The temperature-dependent resistivities of the YbAl$_{3}$, LuAl$_{3}$ + Al, and Fe layers, extracted from the full heterostructures (Supplementary Information), are shown in Fig.~1A.  For spin-torque measurements, the films were patterned using optical lithography and ion milling, and then Ti/Pt electrical contacts were deposited. The samples used for spin-torque ferromagnetic resonance were 6-60 $\mu$m wide and 40-80 $\mu$m long, and the samples for magneto-optical Kerr effect measurements were 50 $\mu$m wide by 60 $\mu$m long. The device geometries are shown in Fig.~2(a) and Fig.~3(a), respectively.


We measured the spin-orbit torques generated by YbAl$_{3}$ and LuAl$_3$ by detecting the deflection of magnetization of the Fe layer in response to charge currents applied in the plane of the heterostructure.  We read out this deflection in two independent ways: electrically via the  spin-torque ferromagnetic resonance (ST-FMR) technique \cite{Mellnik2014, Liu2011} and optically using the magneto-optical Kerr effect (MOKE) \cite{fan2014quantifying}.  We performed ST-FMR on samples with all three thicknesses of YbAl$_3$, observing quantitatively similar behavior.

For ST-FMR, we applied a $f$ = 18-23 GHz oscillating drive current in the presence of a swept external magnetic field at a fixed in-plane angle that orients the equilibrium direction of the Fe magnetization at an angle $\phi$ with respect to the current flow direction.  As the field is swept through the resonance condition of the magnetic layer, the magnetization is driven into ferromagnetic resonance. The resulting precessional motion of the magnetization near resonance gives rise to an oscillating longitudinal resistance through the magnet's anisotropic magnetoresistance (AMR). The mixing of this resistance oscillation and driving signal results in a DC voltage signal, $V_{mix}$, that has the form of a sum of an anti-symmetric and a symmetric Lorentzian.  Figure 2B shows a representative resonance at 40 K for a driving frequency of
22 GHz at $\phi = 45^{\circ}$ with respect to the current flow direction, along with fits.  From the magnitudes of the antisymmetric and the symmetric Lorentzians  we extract the out-of-plane torque ($\tau_{\perp}$) and the in-plane torque ($\tau_{\parallel}$), respectively. By varying the angle at which the external field is applied, we determined the symmetries of the torques that make up $\tau_{\perp}$ and $\tau_{\parallel}$. The component of primary interest is the in-plane antidamping torque $\tau_{\parallel,AD}$, which varies $\propto \cos(\phi_B)$.  For details of the analysis, see section 5.4 in the Methods section.


\begin{figure}[!tbp]
\includegraphics[width=0.8\columnwidth]{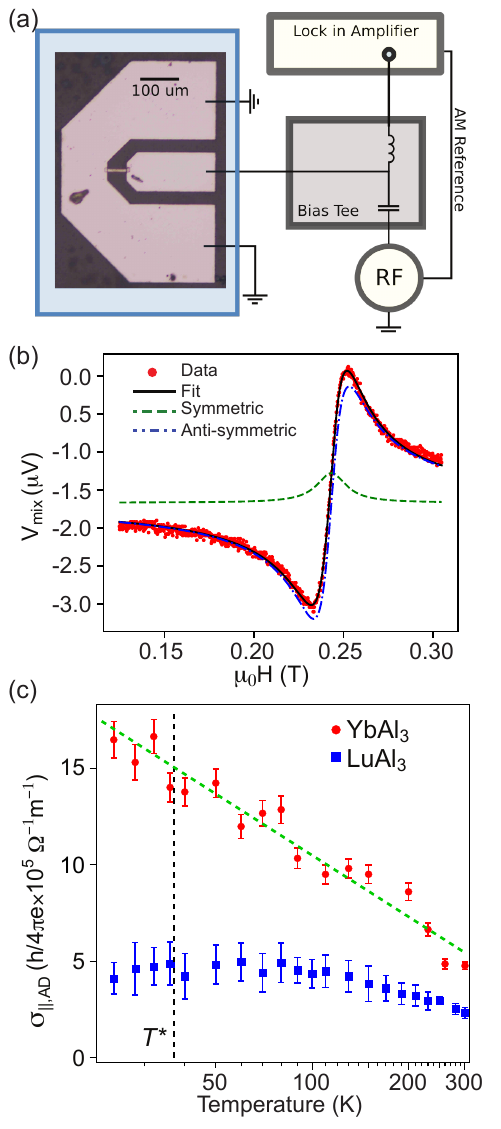}
\centering
   \caption{\textbf{ST-FMR measurements of current-induced torque in a 1.8 nm Al/5 nm LuAl$_3$/6.7 nm YbAl$_3$/6 nm Fe/oxidized Al heterostructure.} (a) Schematic of the ST-FMR measurement setup. (b) Representative DC mixing voltage trace taken at 40 K at a frequency $f$ = 22 GHz. The black line is a fit to Eq.~(S4) (see supplementary information), with the symmetric Lorentzian and antisymmetric Lorentzian components shown in green and blue dot-dash lines, respectively. (c) Spin-orbit-torque conductivity determined via ST-FMR with the external magnetic field at a fixed 45 degree angle. Error bars indicate the standard error. The green dashed line in (c) is a guide to the eye for ln$(\frac{T_{0}}{T})$ temperature scaling.}
    \label{fig2}
\end{figure}

Figure 2(c) displays the results of the ST-FMR measurements for the spin-orbit-torque conductivities $\sigma_{\parallel,AD}$ of YbAl$_3$ and LuAl$_3$, equal to the maximum transverse angular momentum densities absorbed by the magnetic layer per unit applied electric field.  These quantities are related to the internal spin Hall conductivities within the spin-orbit source material, $\sigma_{S}$, as $\sigma_{\parallel,AD}= T_\text{int} \sigma_{S}$ where $T_\text{int} \leq 1$ is an interfacial transmission factor for the spin current to be absorbed by the magnetic layer.

For the control sample in which LuAl$_{3}$ is the spin-orbit source layer, we find that $\sigma_{\parallel,AD,\text{LuAl}_3}$ increases from $(2.3 \pm 0.3)\times10^{5} \frac{\hbar}{2e} \Omega^{-1}$ m$^{-1}$ at 300 K to $(5.0 \pm 1.0)\times10^{5} \frac{\hbar}{2e} \Omega^{-1}$ m$^{-1}$ below 90 K.  This type of weak temperature dependence is similar to previous measurements of antidamping spin-orbit torque generated by the intrinsic spin Hall effect in $d$-electron heavy metals \cite{kim2014anomalous, qiu2014angular, ou2016origin}.The maximum value of the spin-orbit torque ratio in LuAl$_{3}$ is $\xi_{\parallel,AD,\text{LuAl}_3}\approx$ 0.04, where $\xi_{\parallel,AD}= \frac{2e}{\hbar} \frac{\sigma_{\parallel,AD}}{\sigma}$ and $\sigma$ is the charge conductivity in the spin-orbit source layer.   In contrast, the spin-orbit torque from YbAl$_{3}$ has a stronger, more dramatic temperature dependence.  At room temperature, $\sigma_{\parallel,AD,\text{YbAl}_3} = (5.3 \pm 0.5)\times10^{5} \frac{\hbar}{2e} \Omega^{-1}$ m$^{-1}$, comparable to $\sigma_{\parallel,AD,\text{LuAl}_3}$, but with decreasing temperature $\sigma_{\parallel,AD,\text{YbAl}_3}$ increases approximately logarithmically to a value of $(16.1 \pm 1.0)\times10^{5} \frac{\hbar}{2e} \Omega^{-1}$ m$^{-1}$ at the lowest temperature measured, 24 K.  The maximum value of $\sigma_{\parallel,AD}$ for YbAl$_{3}$ is strikingly large -- more than a factor of two larger than for previous measurements of heavy metals (see Table 1 in ref.~\cite{zhu2019strong}).  It is surpassed in the literature only by a small number of topological insulators such as (Bi$_{0.5}$Sb$_{0.5}$)$_2$Te$_3$ and  Bi$_{0.9}$Sb$_{0.1}$ \cite{fan2014magnetization,khang2018conductive}. Even at room temperature the spin Hall conductivity in YbAl$_{3}$ ((5.3 $\pm$ 0.5)$\times$10$^{5}$ $\frac{\hbar}{2e} \Omega^{-1}$ m$^{-1}$) is larger than other well-known heavy metal systems (see Table 1 in ref.~\cite{zhu2019strong})
The sign of $\sigma_{\parallel,AD,\text{YbAl}_3}$ is positive (the same as Pt), as is expected since the 4$f$-shell in Yb is more than half full. 
The spin torque ratio $\xi_{\parallel,AD,\text{YbAl}_3}$ is in the range 0.3-0.4 over the full temperature range, approximately an order of magnitude larger than for the LuAl$_{3}$ control sample.


\begin{figure}
\includegraphics[width=0.8\columnwidth]{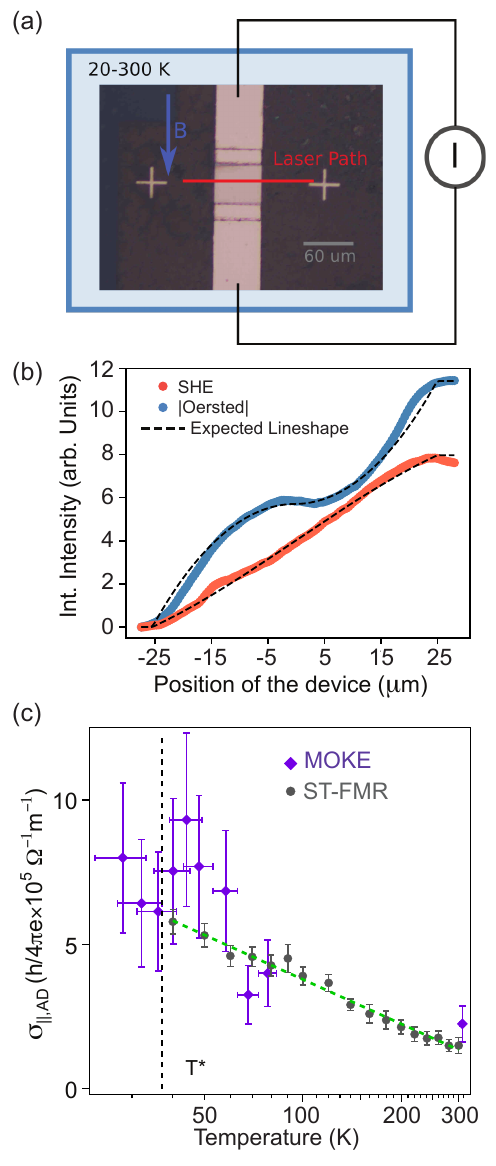}
\centering
   \caption{\textbf{MOKE and angle dependent ST-FMR measurements of current-induced torque in a 1.8 nm Al/5 nm LuAl$_3$/15 nm YbAl$_3$/6 nm Fe/oxidized Al heterostructure.} (a) Path of the laser scan used in the MOKE measurements across a sample 50 $\mu$m wide by 60 $\mu$m long. The laser spot is $\approx$ 10 $\mu$m wide. (b) The cumulative integral of the MOKE signal at 36 K that is  even in magnetization signal (Oersted) and odd in magnetization (SHE). The dashed lines are guides to the eye for the expected lineshapes of the cumulative integrals of the signals. (c) Spin-orbit-torque conductivity determined via MOKE (violet) and angle-dependent ST-FMR (grey). Error bars indicate the standard error. The green dashed line in (c) is a guide to the eye for ln$(\frac{T_{0}}{T})$ temperature scaling.}
    \label{figMoke}
\end{figure}

We note that there is a potential complication that can affect quantitative interpretations of the ST-FMR results. The AMR of Fe is sufficiently weak that the symmetric component of the ST-FMR mixing signal is not necessarily fully dominant over an artifact due to spin pumping and the inverse spin Hall effect (SP/ISHE), which can also induce a symmetric resonance signal\cite{tserkovnyak2002spin, mosendz2010detection, azevedo2011spin}. 
The magnitude of the SP/ISHE signal will depend on several parameters that are poorly-known for our samples, but given the sign of the AMR for Fe, which is opposite to CoFeB, the ST-FMR rectification and SP/ISHE signals will have the same sign\cite{karimeddiny2020transverse}. Thus, if the spin pumping signal is significant this would lead to an overestimate of $\sigma_{\parallel,AD,\text{YbAl}_3}$ above. This complication does not change our primary conclusion about the strong temperature dependence and large magnitude of the spin torque conductivity in YbAl$_{3}$, because SP/ISHE can only affect a resonance signal that grows strongly with decreasing temperature if the spin Hall conductivity itself increases strongly with decreasing temperature (see section V in the supplementary information for a worst-case analysis in which we assume the SP/ISHE signal is dominant over the spin-torque rectification signal). Nevertheless, to check for the  influence of a SP/ISHE signal, we used MOKE to perform independent measurements of the spin-torque conductivity.

The MOKE measurements were modeled after ref.~\cite{fan2014quantifying} and were performed on samples with 15 nm of YbAl$_3$. In the presence of an 80 mT external magnetic field parallel to the current flow direction, we apply a low frequency ($<$6 kHz) AC drive current so that the current-induced torques drive quasi-static equilibrium shifts of the magnetization. To ensure full saturation of the Fe layer, we use a bar fabricated along the Fe [100] direction easy axis (Fe[100]$||$YbAl$_3$ [110]). Using a polar MOKE geometry so that we are sensitive only to the out-of-plane component of the magnetization, and 45$^\circ$ linearly polarized light to minimize quadratic MOKE contributions (0$^\circ$ corresponds to fully $s$ polarized), we scan our laser spot across a 60 $\mu$m wide by 50 $\mu$m long device perpendicular to the current flow direction, and read out the spatial variation of the magnetization oscillation.  We perform the same experiment at opposite magnetic fields to separate the Kerr response signals into odd (i.e., due to the effective field from the antidamping spin-orbit torque) and even (i.e., Oersted field) functions of magnetization.  The MOKE signal is calibrated by comparing the position-dependent part of the even signal to the easily-calculable out-of-plane component of the Oersted field (Fig.~3(b), see section VI in the supplementary information). The temperature stability of our optical cryostat limits the MOKE results to room temperature and below 77 K. However in this range the MOKE values for the antidamping spin-orbit torque conductivity confirm the very large values of $\sigma_{\parallel,\mathrm{AD},\text{YbAl}_3}$ indicated by ST-FMR at low temperature (Fig.~3(c)). The difference in magnitude of $\sigma_{\parallel,AD,\text{YbAl}_3}$ between Fig.~\ref{fig2} (6.7 nm thick YbAl$_3$) and Fig.~\ref{figMoke} (15 nm thick YbAl$_3$) may be due to differences in interface quality between the two growths. 

\begin{figure}
\includegraphics[width=1\columnwidth]{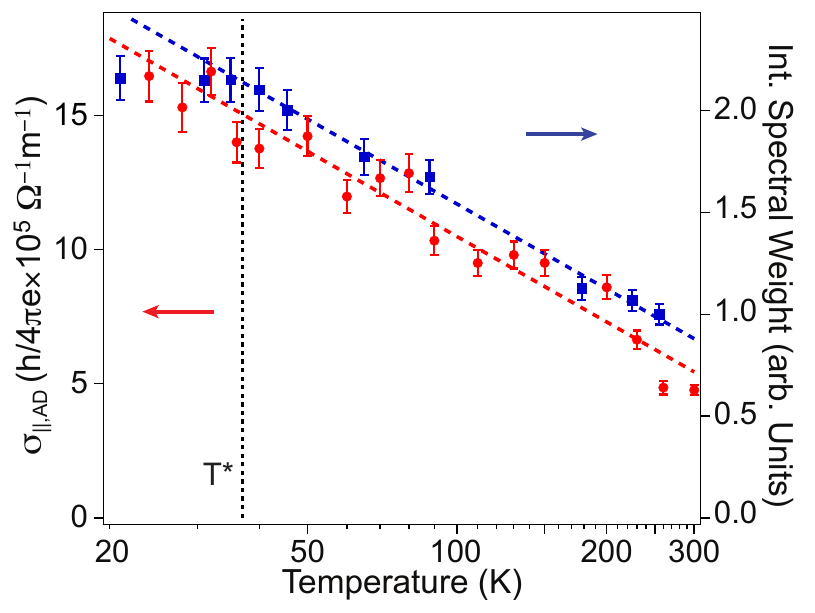}
\centering
   \caption{\textbf{Scaling comparison of  $\sigma_{||,\mathrm{AD},\text{YbAl}_3=6.7\text{ nm}}$ and YbAl$_3$ 4$f$ spectral weight with temperature}. $\sigma_{||,\mathrm{AD}}$ (red circles, left axis, error bars are the standard error) is compared to the integrated 4$f$ spectral weight (blue squares, right axis, error bars are 3\% margin of error). The vertical dashed line highlights the coherence temperature $T^*$=37 K of YbAl$_3$. The colored, parallel dashed lines are guides to the eye $\propto \ln\left(\frac{T_0}{T}\right)$.}
    \label{fig3}
\end{figure}

A comparison between the temperature dependence of the integrated spectral weight measured by ARPES and the spin torque conductivity measured by ST-FMR  for the 6.7 nm YbAl$_3$ sample, $\sigma_{\parallel,AD}$, is shown in Fig.~\ref{fig3}.  Both quantities increase with decreasing temperature with approximately a log$\frac{T_{0}}{T}$ scaling.  The measured enhancement in the integrated spectral weight ranges between a factor of approximately 2 and 4 depending on the integration window and the background subtraction (see section VII in the supplementary information), very similar to the factor of 3 enhancement observed in $\sigma_{\parallel,AD}$. 

The logarithmic temperature dependence for the antidamping spin-orbit-torque conductivity of YbAl$_3$, an enhancement of $\sigma_{\parallel,AD}$ by at least a factor of 3 compared to the room temperature value, is quite unusual.  In $d$-electron heavy metals, the spin-orbit torque varies at most about 10\% from room to cryogenic temperatures \cite{qiu2014angular, ou2016origin}.  As far as we are aware, the only strong temperature dependencies observed previously for $\sigma_{\parallel,AD}$ in spin-orbit-torque metals are in ferromagnetic alloys, which can exhibit approximately a factor of 4 enhancement near their Curie temperature due to scattering from magnetic fluctuations \cite{ou2018strong} and in high-resistivity SrRuO$_{3}$\, which shows a factor of 8 enhancement with decreasing temperature associated with the variation in carrier lifetime in the bad-metal regime \cite{ou2019exceptionally, Tanaka2008}. Neither of these mechanisms is applicable to YbAl$_{3}$, since it does not undergo a magnetic phase transition and it has a low resistivity ($< 50$ $\mu\Omega$-cm, see Fig.~1(c)) over the entire temperature range we consider. 

Thus, in light of the ARPES measurements which have identified the temperature-dependent changes in the electronic structure of YbAl$_{3}$ as due to increasing effective hybridization between Yb 4$f$ local moments and delocalized conduction electrons, resulting in the formation of a renormalized electron bandstructure near the Fermi level, and based on the similarities in the logarithmic temperature scaling, we identify the large enhancement in spin-torque conductivity as due to the same many-body electron correlations. Our observations are qualitatively consistent with predictions in ref.~\cite{Reynolds2017} that the enhancement in spin-orbit torque can be achieved by increasing hybridization of the 4$f$ states into the itinerant bands near the Fermi level.

The strong enhancement that we measure of the spin-orbit torque associated with the Kondo effect in YbAl$_{3}$  suggests that engineering of many-body states in correlated electron systems (particular those containing partially filled $f$ shells) can be a productive strategy for increasing the efficiency of spin-orbit torques. The same sensitivity may also enable spin-orbit torques to be a useful new tool for characterizing heavy fermion systems.

 Primary support for this work was provided by the National Science Foundation (DMR-1708499 and DMR-1709255), and by the Gordon and Betty Moore Foundation as part of the EPiQS initiative (GBMF3850).  Additional support was provided by the NSF MRSEC program through the Cornell Center for Materials Research (DMR-1719875). This work was performed in part at the Cornell NanoScale Facility, an NNCI member supported by the NSF (NNCI-1542081) and at the Cornell Center for Materials Research shared facilities.

%



\end{document}


\title{Strongly Temperature-Dependent Spin-Orbit Torques in Heavy Fermion YbAl$_3$}
\author{Neal D. Reynolds} 
\email{ndr37@cornell.edu}
\thanks{equal contribution}
\affiliation{Department of Physics, Cornell University, Ithaca, NY 14853, USA}
\author{Shouvik Chatterjee}
\email{sc2246@cornell.edu}
\thanks{equal contribution}
\affiliation{Department of Physics, Cornell University, Ithaca, NY 14853, USA}
\author{Gregory M. Stiehl}
\affiliation{Department of Physics, Cornell University, Ithaca, NY 14853, USA}
\author{Joseph A. Mittelstaedt}
\affiliation{Department of Physics, Cornell University, Ithaca, NY 14853, USA}
\author{Saba Karimeddiny}
\affiliation{Department of Physics, Cornell University, Ithaca, NY 14853, USA}
\author{Alexander J. Buser}
\affiliation{Department of Physics, Cornell University, Ithaca, NY 14853, USA}
\author{Darrell G. Schlom}
\affiliation{Department of Materials Science and Engineering, Cornell University, Ithaca, NY 14853, USA}
\affiliation{Kavli Institute at Cornell, Cornell University, Ithaca, NY 14853, USA}
\affiliation{Leibniz-Institut f\"{u}r Kristallz\"{u}chtung, Max-Born-Str. 2, 12489 Berlin, Germany}
\author{Kyle M. Shen}
\affiliation{Department of Physics, Cornell University, Ithaca, NY 14853, USA}
\affiliation{Kavli Institute at Cornell, Cornell University, Ithaca, NY 14853, USA}
\author{Daniel C. Ralph}
\affiliation{Department of Physics, Cornell University, Ithaca, NY 14853, USA}
\affiliation{Kavli Institute at Cornell, Cornell University, Ithaca, NY 14853, USA}



\maketitle 

\section{Thin Film Growth}
Single crystalline, epitaxial YbAl$_{3}$ thin films were synthesized on MgO (001) substrates using a Veeco Gen10 molecular-beam epitaxy (MBE) system with base pressure better than 2$\times$10$^{-9}$ Torr. Prior to growth, the MgO substrates were annealed in vacuum for 20 min at 800$^{\circ}$C. A 1.8 nm thick aluminum (Al) buffer layer was deposited at 500$^\circ$C followed by a 5 nm thick LuAl$_{3}$ layer on which YbAl$_{3}$ layers were synthesized. Al and LuAl$_{3}$ buffer layers were found to be essential for the synthesis of high quality YbAl$_{3}$ thin films. For the growth of LuAl$_{3}$ and YbAl$_{3}$ layers, Lu/Yb and Al were co-evaporated from effusion cells at a rate of $\approx$ 0.4 nm/min onto a rotating substrate  with real-time reflection high-energy electron diffraction (RHEED) monitoring.  The LuAl$_{3}$ atomic layers were deposited at 200$^\circ$C and annealed at 350$^\circ$C for 30 min, following which YbAl$_{3}$ deposition was initiated at 200$^\circ$C and ramped up to 315$^\circ$C during growth. Due to the co-evaporation method, the surface termination was not deliberately controlled. LuAl$_3$ and YbAl$_3$ films for ARPES measurements were measured immediately \textit{in situ} without breaking vacuum.  For the spin-torque samples, Fe layers were deposited either on LuAl$_{3}$ or on YbAl$_{3}$ layers at 200$^{\circ}$C followed by the deposition of a 3 nm thick Al capping layer at room temperature to prevent oxidation of the underlying heterostructure. Further details about the thin film growth and characterization of the heterostructure can be found in the supplementary information and in ref.~\cite{chatterjee2016epitaxial}

\section{Resistivity determination for the layers in the LuAl$_3$ / YbAl$_3$ /Fe / AlO$_x$ stack}

\begin{figure*}
\includegraphics[width=.5\textwidth]{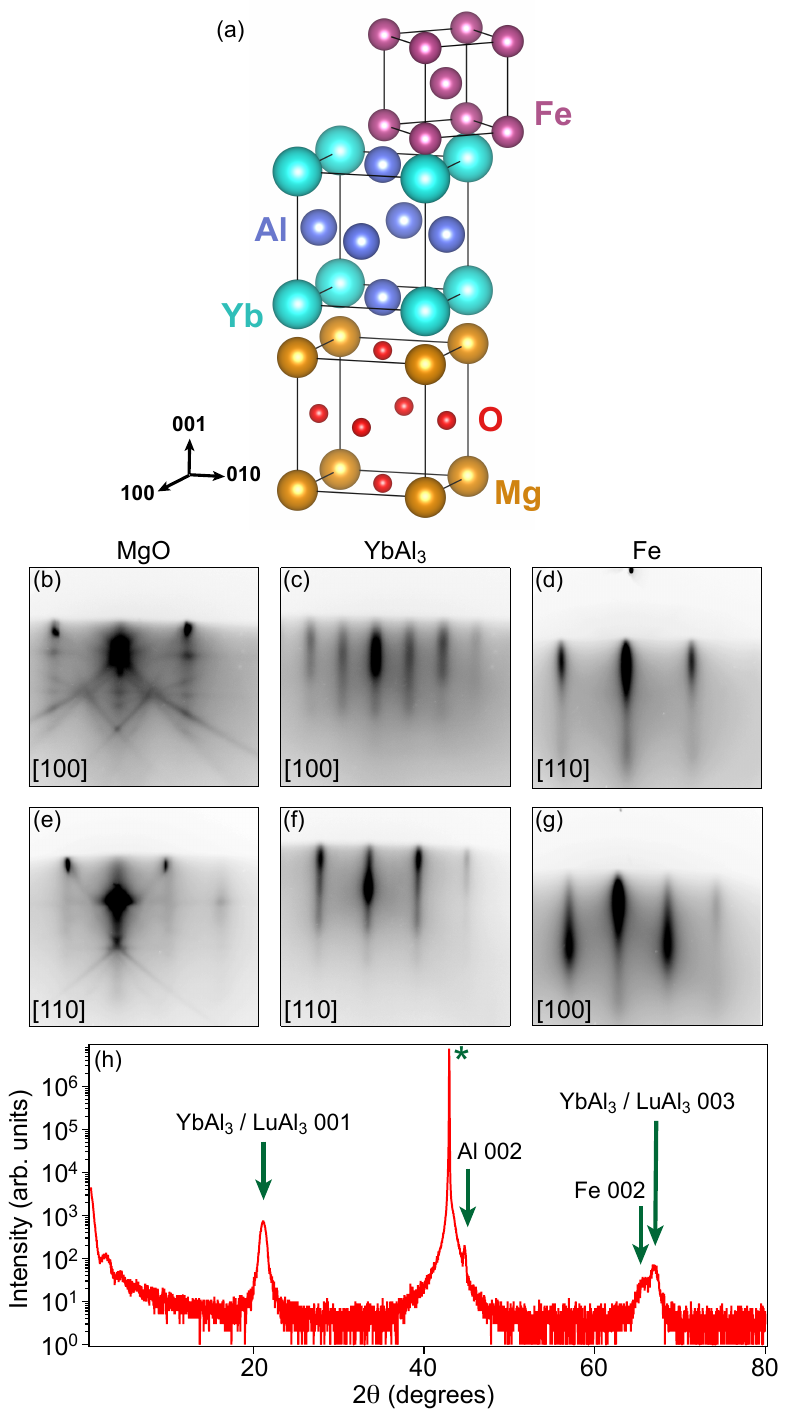}
\centering
   \caption{\textbf{Structural characterization and orientation of YbAl$_3$ and Fe layers.} (a) An atomic model illustrating the orientation relationship between MgO, YbAl$_{3}$ and Fe layers in the heterostructure used for ST-FMR and MOKE measurements. (b-g) Reflection high-energy electron diffraction (RHEED) images at different stages of the growth process of a heterostructure similar to the one used for ST-FMR and MOKE measurements. Images taken along the [100] azimuth of the MgO substrate (b) prior to the start of the growth process, (c) after YbAl$_{3}$ deposition, and (d) after Fe deposition. (e-g) Corresponding RHEED images along the [110] azimuth of the MgO substrate. (h) Out-of-plane $\theta$ - 2$\theta$ x-ray diffraction scan (XRD) of the heterostructure taken with Cu K$\alpha_{1}$ radiation. The substrate peak is indicated by an asterisk. The RHEED images along with the XRD scan establish the epitaxial relationship between MgO, YbAl$_{3}$ and Fe layers in the heterostructure.}
    \label{fig:RHEED}
\end{figure*}

We grew a series of heterostructures with different thicknesses for the YbAl$_3$ and Fe layers: Al 1.8 nm/LuAl$_3$ 5 nm/YbAl$_3$ \textit{x} nm/Fe 6 nm/Al 3 nm (\textit{x} = 6.7,15,18) and Al 1.8nm/LuAl$_3$ 5 nm/YbAl$_3$ 10 nm/Fe \textit{y} nm/Al 3 nm (\textit{y} = 3,6,9,12). In the heterostructures the epitaxial orientation relationship among Fe and YbAl$_{3}$/LuAl$_{3}$ with respect to MgO is (001) [110] Fe $\vert\vert$ (001) [100] YbAl$_{3}$/LuAl$_{3} \vert\vert$ (001) [100] MgO (Fig. \ref{fig:RHEED}). We assume that the top Al layer oxidizes completely upon exposure to air and thus does not contribute to the resistance of the stacks. Hall bars with length $\times$ width dimensions of 560 $\mu$m $\times$ 110 $\mu$m, 240 $\mu$m $\times$ 40 $\mu$m, and 120 $\mu$m $\times$ 20 $\mu$m were fabricated on the same die as the magneto optical Kerr effect (MOKE) devices and spin torque ferromagnetic resonance (ST-FMR) devices. For each film, the temperature-dependent resistivity was measured using a Hall-bar device in a Quantum Design physical properties measurement system (PPMS). We find excellent agreement between sheet resistances of the Hall bars on the same die as a function of temperature.  
We extract the resistivities $\rho_{\text{YbAl}_3}$ and $\rho_{\text{Fe}}$ from the slope of the linear regression of the sheet resistance vs.\ 1/\textit{x} or 1/\textit{y}, respectively. We determine $\rho_{\text{LuAl}_3+\text{Al}}$ from the intercept of the sheet resistance vs. 1/\textit{x} (the YbAl$_3$ thickness series) after subtracting out the contribution due to the Fe layer using $\rho_{\text{Fe}}$ determined from the thickness series of Fe. The resulting $\rho$ vs. T for each layer is shown in Fig. 1(c) of the main text. $\rho_{\text{YbAl}_3}$ and $\rho_{\text{LuAl}_3+\text{Al}}$ are consistent with our previous work\cite{chatterjee2016epitaxial} and $\rho_{\text{Fe}}$ is consistent with earlier reports as well\cite{taylor1968resistivity}.

\section{Magnetic characterization of the Fe layer}

\begin{figure*}
\includegraphics[width=1\textwidth]{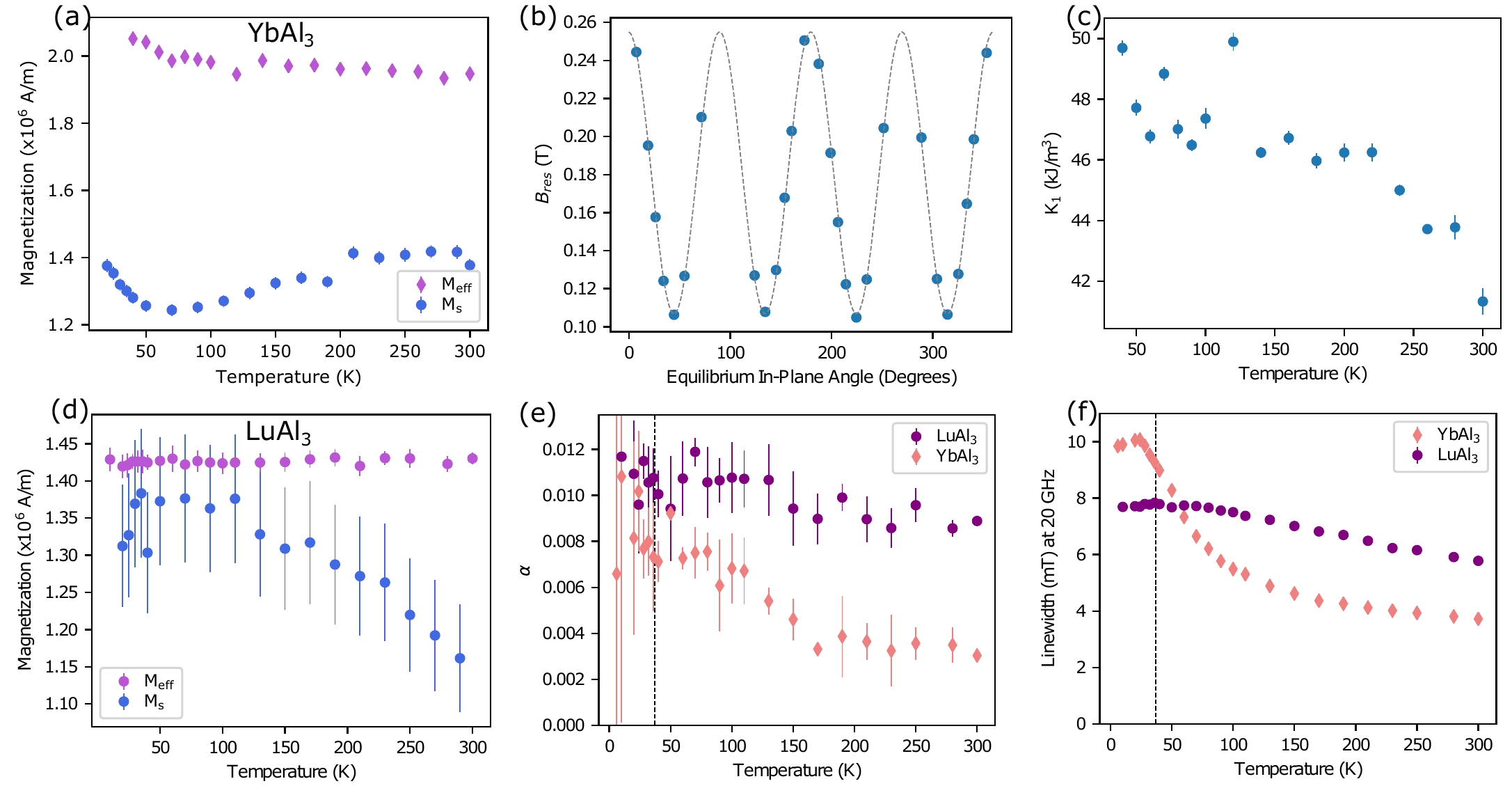}
\centering
   \caption{\textbf{Magnetic properties of Fe layer grown on YbAl$_3$ and LuAl$_3$.} Magnetic characterization of the Fe layer in both YbAl$_3$/Fe (i.e., MgO (001)/Al 1.8 nm/LuAl$_3$ 5 nm/YbAl$_3$ 15 nm/Fe 6 nm/oxidized Al) and LuAl$_3$/Fe (i.e., MgO (001)/Al 1.8 nm/LuAl$_3$ 15 nm/Fe 6 nm/oxidized Al). (a,d) Fe effective magnetization derived from the resonance condition in ST-FMR (Eq.~(\ref{eq:tetraMAE_res}) and the saturation magnetization extracted from vibrating sample magnetometry (purple diamonds and blue circles, respectively) of the (a) YbAl$_3$/Fe and (d) LuAl$_3$/Fe. (b) The ST-FMR resonant field as a function of in-plane magnetization angle for the YbAl$_3$/Fe structure at 40 K (blue circles) with a fit to Eq.~(\ref{eq:tetraMAE_bres}) (dashed line). (c) Magnitude of the cubic anisotropy constant, K$_1$, extracted from the measurements in (b). (e) Gilbert damping parameter derived from a linear fit to the linewidth of the ST-FMR resonance curves vs.\ frequency from 18-20 GHz for YbAl$_3$/Fe (orange diamonds) and LuAl$_3$/Fe (purple circles), for magnetic fields applied along a magnetic easy axis of the Fe layer. (f) Linewidth at 20 GHz vs.\ temperature for YbAl$_3$/Fe (orange diamonds) and LuAl$_3$/Fe (purple circles). The vertical dashed line in (e) and (f) is at 37 K, the temperature below which Fermi liquid behavior is observed. }
    \label{fig:SImagnets}
\end{figure*}

As discussed in the Methods section of the main text, we perform vibrating sample magnetometry (VSM) using a Quantum Design PPMS to extract the sample saturation magnetization, $M_s$, for both the Fe layer in the YbAl$_3$/Fe heterostructure (i.e. MgO (001)/Al 1.8 nm/LuAl$_3$ 5 nm/YbAl$_3$ 15 nm/Fe 6 nm/oxidized Al) and the Fe layer in the LuAl$_3$-only control heterostructure (i.e. MgO (001)/Al 1.8 nm/LuAl$_3$ 15 nm/Fe 6 nm/oxidized Al), yielding the $M_s$ vs.\ T dependence shown in Fig.~\ref{fig:SImagnets}(a) and (d), respectively. The temperature dependence of the magnetization in the YbAl$_3$/Fe structure appears qualitatively to follow the YbAl$_3$ susceptibility vs.\ T (see ref.~\cite{hiess1995magnetic}) in reverse, \textit{i.e.} $M_s$ decreases as $\chi_{\text{YbAl}_3}$ increases, suggesting a proximity effect in which a small induced moment in YbAl$_3$ is antiparallel to that in the Fe layer. In both the YbAl$_3$ stack and the LuAl$_3$ only stack, the Fe layer saturation magnetization is slightly smaller than the 1.71$\times$10$^6$ A/m saturation magnetization for bulk, crystalline Fe \cite{crangle1971magnetization}.

The in-plane angular dependence of the resonant field reveals a moderate bi-axial magnetocrystalline anistropy field with easy axes along the substrate diagonals, corresponding to the Fe [100] directions (and parallel to the YbAl$_3$ [110] directions, see Fig.~\ref{fig:RHEED}). No additional uniaxial anisotropy was observed, which is expected because the substrates were continuously rotated during the deposition process. A representative trace at 40 K is shown in Fig.~\ref{fig:SImagnets}(b)) with a fit to 
\begin{multline}
    B_{res}=\frac{1}{8}\left(\sqrt{64\frac{\omega^2}{\gamma^2}+9H_a^2+24H_a M_{\mathrm{eff}}+16M_{\mathrm{eff}}^2+9H_a^2\cos^2(4\phi)+(18H_a^2+24H_a M_{\mathrm{eff}})\cos(4\phi)}\right.\\
    \left.+5H_a\cos(4\phi)-3H_a-4M_{\mathrm{eff}} \right) \label{eq:tetraMAE_bres}
\end{multline}
where $B_{res}$ is the resonant field, $\phi$ is the equilibrium direction of the magnetization for a given in-plane field angle with respect to the YbAl$_{3}$ [100] direction, $\omega$ is the angular frequency of the driving current, $\gamma$ is the electron gyromagnetic ratio, $H_a$ is the anisotropy field defined as $2K_1/M_s$ (where $K_1$ is the first order cubic anistropy constant), and $M_{\mathrm{eff}}$ is the effective magnetization including any contributions from an out of plane uniaxial anisotropy.  Eq.~\ref{eq:tetraMAE_bres} is the solution for $B_{res}$ derived from resonance condition arising from an in-plane magnetized film with tetragonal magnetocrystalline anisotropy\cite{farle1998ferromagnetic}:
\begin{equation}
    f=\frac{\gamma}{2\pi}\sqrt{\left(B_{res}-2H_a\cos(4\phi)\right)\left(B_{res}+\mu_0M_{\mathrm{eff}}+\frac{H_a}{2}(3-\cos(4\phi))\right)} \label{eq:tetraMAE_res}
\end{equation}
The fit is recursively solved with the value of the equilibrium direction of the magnetization for each applied field angle recomputed at each step in the fitting process. This yields a K$_1$ of 41 kJ/m$^3$ at 300 K which increases as the temperature is lowered to 50 kJ/m$^3$ at 40 K   (Fig.~\ref{fig:SImagnets}(c)). As with M$_s$, these values for K$_1$ are slightly smaller than the K$_1$ found in crystalline Fe (48 kJ/m$^3$ at 300 K, 52 kJ/m$^3$ at 77 K) \cite{graham1958magnetocrystalline}.

From our fits to the resonant field vs.\ angle and at each temperature we also extract the effective magnetization shown as the purple diamonds in Figs.~\ref{fig:SImagnets}(a) and (d). We find an $M_{\mathrm{eff}}$ that increases with decreasing temperature and is non-trivially larger than $M_s$ at a given temperature. This suggests that the small anti-parallel moment induced in the YbAl$_3$ layer also stabilizes the magnetization in-plane.

Figs. S2(e) and S2(f) plot the Gilbert damping parameter, $\alpha$, vs.\ temperature. $\alpha$ is extracted from the  the frequency dependence of the linewidth, $\Delta$ (determined from the fits to Eq.S4, described in section IV, to our ST-FMR experiments), via a linear fit:    
\begin{equation}
    \Delta = \frac{\alpha\omega}{\gamma}+\Delta_{\mathrm{inhomo}},\label{eq:gilbertdamping}
\end{equation}
where $\omega$ is the angular frequency of the applied RF current, $\gamma$ is the electron gyromagnetic ratio, and $\Delta_{\mathrm{inhomo}}$ is the inhomgenous component of the linewidth. Fitting to $\Delta$ vs $\omega$ for $\frac{\omega}{2\pi}$=18-20 GHz yields the values of $\alpha$ shown in Fig. S2(e). We find the value of  $\Delta_{\mathrm{inhomo}}$ to be negligible to within our error. The room temperature damping of $\approx$0.003 is consistent with the reported range of 0.002-0.004.\cite{fedamping}  The peak in the linewidth vs temperature observed in Fig. S2(f) is characteristic of rare earth impurities in the ferromagnetic Fe layer,\cite{BaileyPyRedoped,PhysRev.133.A728, reidy2003dopants} and is likely a consequence of the direct growth of the Fe layer on the YbAl$_{3}$.

\section{Spin-Torque Ferromagnetic Resonance (ST-FMR)}

We perform ST-FMR measurements using a radio frequency (RF) insert within a sample-in-vapor He flow cryostat. A semi-rigid coaxial line is terminated in a coplanar waveguide to which we wirebond our device in a ground-signal-ground configuration. We apply a 18-23 GHz RF current to the device in the presence of a swept magnetic field at an angle $\phi_B$ with respect to the current flow direction. For the data shown in the main text, current flows along the YbAl$_3$ [100] direction. The magnetic field is provided by an external electromagnet mounted on a motorized base. Due to the relatively strong tetragonal magnetic anisotropy, the magnetization angle $\phi$ is generally slightly misaligned with $\phi_B$.  We perform fits relative to the actual magnetization angle $\phi$ rather than $\phi_B$, calculating $\phi$ based on the measured anisotropy and the applied field strength. Current-driven precession of the magnetization results in an oscillation of the longitudinal resistance of the device through the Fe layer's anisotropic magnetoresistance. The in-phase component of this resistance oscillation mixes with the RF current to create a DC voltage $V_{mix}$ across the device given within a macrospin approximation by \cite{Liu2011}:
\begin{equation}
V_{mix}\approx\frac{(R_{\parallel}-R_{\perp})\sin(\phi)\cos(\phi)I_{0}\omega\tau_{0}}{\gamma^{2} \Delta (2B_{0}+\mu_{0} M_{\mathrm{eff}})}\times\\
\left[S(\phi)F_{S}(B)+\sqrt{1+\frac{\mu_{0} M_{\mathrm{eff}}}{B_{0}}}A(\phi)F_{A}(B)\right]\label{eq:stfmr}
\end{equation}
where $(R_{\parallel}-R_{\perp})$ is the difference in device resistance with the magnetization parallel and perpendicular to the current flow direction; $I_{0}$ is the RF current reaching the device; $\omega$ is the frequency of the driving current; $\tau_{0}\equiv\gamma\frac{\hbar}{2e}\frac{J_{c,YbAl_{3}}}{M{_{s}}t{_{mag}}}$; $J_{c,YbAl_{3}}$ is the charge current density flowing within the YbAl$_{3}$ layer; $M_{s}$ is the saturation magnetization of the Fe layer; $t_{mag}$ is the thickness of the magnetic layer; $\Delta$ is the resonance linewidth as a function of magnetic field; $B_{0}$ is the resonant field; $M_{\mathrm{eff}}$ is the effective magnetization; $F_{S}(B)=\frac{\Delta^2}{(B-B_{0})^2+\Delta^2}$ is a symmetric Lorentzian; and $F_{A}(B)=\frac{B-B_{0}}{\Delta}F_{S}(B)$ is an antisymmetric Lorentzian. 

We fit the amplitude of the symmetric component, containing information on the in-plane (IP) torques, to the form
\begin{equation}
 S(\phi)\equiv\cos(\phi)\xi_{\parallel,AD} +\xi_{\parallel,0}.
\end{equation}
The term $\xi_{\parallel,0}$ is included to account for any torque due to an out-of-plane Oersted field that can result from unequal current flow in the two branches of the ground-signal-ground contact geometry (e.g., due to different wirebond impedance; this term is always small relative to $\xi_{\parallel,AD}$). We fit the amplitude of the anti-symmetric Lorentzian, containing information on the out-of-plane  (OOP) torques to the form
\begin{equation}
    A(\phi)\equiv\cos(\phi)(\xi_{\perp,FL}+\xi_{Oe}),
\end{equation}
accounting for contributions from an OOP field-like spin-orbit torque and the torque due to the in-plane Oersted field generated by current flowing within the plane of the heterostructure:
\begin{equation}
    \xi_{Oe}\equiv\frac{\gamma\mu_{0}(J_{c,YbAl_{3}} t_{Yb_{3}}+J_{c,LuAl_{3}+Al}t_{LuAl_{3}+Al})}{2\tau_{0}}. \label{eq:xioe}
\end{equation}
Note that this definition accounts for the fact that current density in the LuAl$_3$ and Al seed layers contribute to the Oersted field torque, despite the fact that Eq.~(\ref{eq:stfmr}) is defined in terms of the current density through the YbAl$_3$ layer only. The antidamping spin-orbit torque conductivity is $\sigma_{\parallel,AD}=\xi_{\parallel,AD} \times \sigma$, where $\sigma$ is the charge conductivity of the YbAl$_{3}$ layer. 

\begin{figure*}
\includegraphics[width=0.75\textwidth]{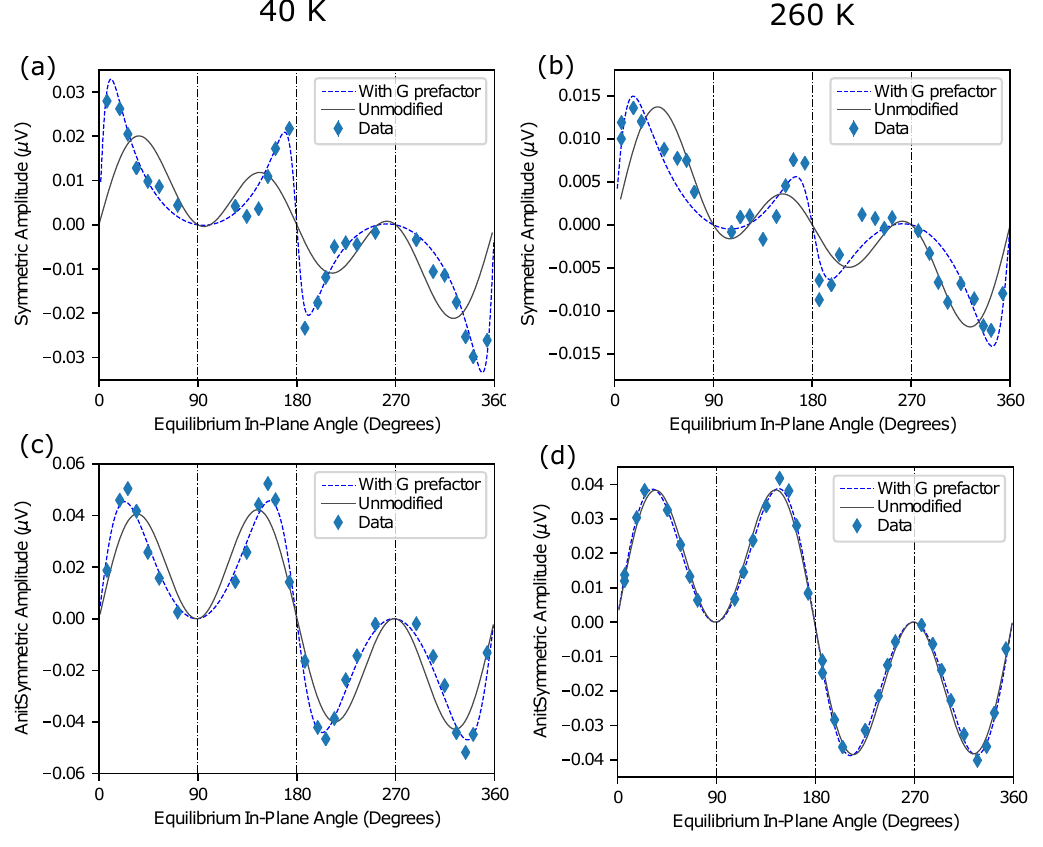}
\centering
   \caption{\textbf{Angle ST-FMR analysis of Fe/YbAl$_3$ heterostructure.} (blue diamonds) Amplitudes of the (a,b) symmetric and (c,d) antisymmetric ST-FMR signals at $f$ = 21 GHz as a function of the angle of in-plane magnetization for the YbAl$_3$/Fe structure at (a,c) 40 K  and (b,d) 260 K. The solid grey lines are fits to Eq.~(\ref{eq:fullanglestfmr}) while the dashed blue lines are the fits to Eq.~(\ref{eq:correctedanglestfmr}) (i.e., the ST-FMR angular dependence with the phenomenological correction factor, G). The vertical dashed lines correspond to the Fe layer magnetic hard axes (Fe [110]). The strength of the G term, and thus the degree to which the signal magnitudes grow near $\phi = 0^{\circ}, 180^{\circ}$, decreases at higher temperatures. }
    \label{fig:SI3}
\end{figure*}

To fully analyze our ST-FMR data, we perform two different measurements. In the first, we apply a fixed frequency (21 GHz) and at each temperature measure the ST-FMR signal versus applied magnetic field magnitude for different angles of the applied  field with respect to the current flow direction. From these scans, we extract the amplitude of the symmetric and antisymmetric Lorentzian components (see eqn. S4) vs.\ the equilibrium in-plane magnetization angle, $\phi$ (where $\phi$=0 corresponds to the YbAl$_{3}$ [100] direction, the Fe [110] direction, and the device current flow direction, see above). Representative angular dependencies are shown in Fig.~\ref{fig:SI3}. The most general angular dependence allowed within a macrospin model for the  symmetric and antisymmetric amplitudes of an ST-FMR resonance due to torque from a nonmagnetic material like YbAl$_3$ has the functional form:
\begin{equation}
F(\phi)=\cos(\phi)\sin(\phi)(R \cos(\phi)+D \sin(\phi)+Z) \label{eq:fullanglestfmr}
\end{equation}
with $R$, $D$, and $Z$ potential components with Rashba-like, Dresselhaus-like, and out-of-plane uniaxial symmetry, respectively. The Z term for the antisymmetric component and the Dresselhaus-like terms for both components are not symmetry-allowed for YbAl$_3$/Fe samples, but we include them in fits as a check for artifacts. Even with this full angular dependence, we can observe deviations from the expected behavior in the regions just above and below $\phi = 0^{\circ}$ and $180^{\circ}$, the two magnetic hard axes where the current-induced torques are strong (see Fig.~\ref{fig:SI3}(a,b,d)).  Because these deviations have an angular dependence incompatible with any component of current-induced torque in the macrospin model, we ascribe them to departures from macrospin dynamics.  Iron has a relatively large value of $M_{\mathrm{eff}}$ and our apparatus is limited to  frequencies $f <$ 23 GHz so as a consequence we are limited to measuring resonances with relatively small values of $B_{res}$ for which the Fe magnetization may not remain fully saturated near the hard axes.  To obtain quantitative results, we can determine the Rashba-like torques using only the ST-FMR amplitudes near the easy axes  $\phi = \pm 45^{\circ}, \pm 135^{\circ}$.  Alternately, we find that we obtain good fits to the full angular dependence by including a phenomenological angle-dependent scaling factor to account for an increased detection sensitivity near $\phi = 0^{\circ}, 180^{\circ}$:
\begin{equation}
F_{corr}(\phi)=\frac{\cos(\phi)\sin(\phi)}{1+G\cos(2\phi)}(R \cos(\phi)+D \sin(\phi)+Z), \label{eq:correctedanglestfmr}
\end{equation}
with $G$ a fitting parameter.  (Note that the added scaling factor does not change the values from $F(\phi)$ at $\phi = \pm 45^{\circ}, \pm 135^{\circ}$, so fit values obtained are consistent with fixed-angle analyses at these angles.) Using this fitting function, as expected we observe large values only for the in-plane and out-of-plane torques with Rashba-like symmetries, with smaller contributions  in some samples for the in-plane torque with out-of-plane uniaxial symmetry (at most ~10 \% of the in-plane torque with Rashba-like symmetry at low temperature where the in-plane torque is large).  We suspect that the term with uniaxial symmetry is due to an out-of-plane Oersted field resulting from unequal current flow in the two arms of the ground-signal-ground contact geometry.

To confirm the absolute magnitude of ST-FMR results, we perform a second experiment to determine the microwave current reaching the device, $I_{0}$.  The wirebonds used to make contact to the device make it difficult to measure the microwave transmission coefficients accurately using a vector network analyzer. Instead, we utilize a variation on the thermal power calibration method described in ref.~\cite{tshitoyan2015electrical}. We measure the device resistance, R, as a function of applied DC bias, I$_{DC}$, from -10 mA to 10 mA and confirm the expected I$_{DC}^2$ dependence of R. We then apply 0 dBm to 20 dBm of RF power at a given frequency to the device in addition to a 500 $\mu$A DC sensing current to the device to extract the RF power dependence of R. Because all of the DC bias current reaches the device, we can then compare the slopes of the device resistance vs total power applied to the device for both the DC bias and RF bias to extract the fraction of RF power which reaches the device. We do this for frequencies from 15-23 GHz at temperature from 20-300 K to map out the scattering parameters of our cryostat and device thus allowing us to calculate $I_0$ for our devices. To verify the accuracy of this analysis, we compare our thermal power calibration scattering parameters to vector network analyzer derived values on our probe station which is not confounded by the problem of wirebonds.

\begin{figure*}
\includegraphics[width=.8\textwidth]{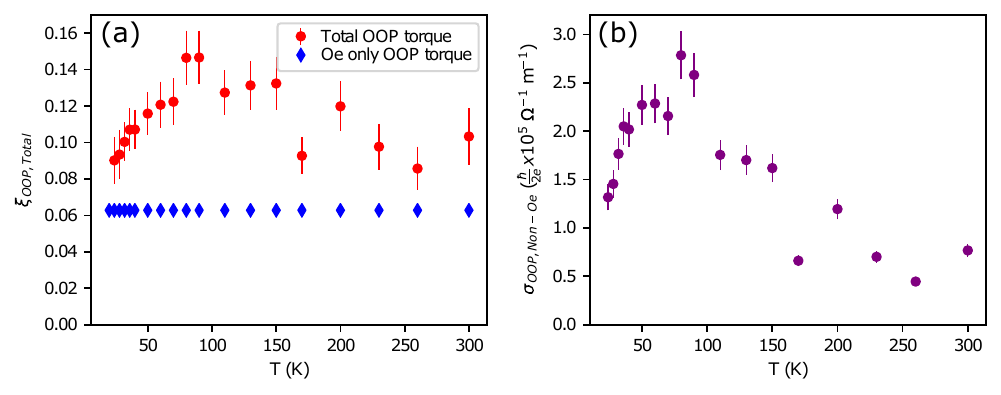}
\centering
   \caption{\textbf{Comparison of measured out-of-plane torque efficiency to the predicted out-of-plane torque efficiency from the Oersted field alone.} (a) (red circles) Measured out-of-plane torque efficiency at $f$ = 23 GHz in the 1.8 nm Al/5 nm LuAl$_3$/15 nm YbAl$_3$/6 nm Fe/ oxidized Al heterostructure as a function of temperature. (blue diamonds) The calculated out-of-plane torque efficiency due to the Oersted field alone. (b) Effective spin-torque conductivity of the out-of-plane field-like spin-orbit torque determined from the difference between the measured changes and the expected Oersted contribution in (a). }
    \label{fig:SI4}
\end{figure*}

The above calibration allows us to quantitatively determine the magnitude of the spin-orbit torques. Comparing the expected out-of-plane torque due the Oersted field with the measured out-of-plane calculated using the calibrated RF current reveals the presence of non-Oersted out-of-plane field-like torque as shown in Fig.~\ref{fig:SI4}(a). The extracted spin torque conductivity for the non-Oersted out-of-plane field-like torque is shown in Fig.~\ref{fig:SI4}(b).  Due to this non-Oersted field-like torque, it is not accurate to apply
the simple ``S/A'' analysis initially suggested in ref.~\cite{Liu2011}.

\section{Spin Pumping Analysis}

The precession of the magnetization in our ST-FMR measurements leads to the generation of a spin current flowing from the magnetic Fe layer in YbAl$_3$ layer with spin direction parallel to the equilibrium magnetization direction\cite{tserkovnyak2002spin}. The spin current can be then rectified into a voltage via the inverse spin Hall effect\cite{mosendz2010detection, azevedo2011spin} in the YbAl$_3$.  This spin-pumping/inverse spin Hall effect (SP/ISHE) signal is difficult to disentangle from the rectification signal due to antidamping spin-orbit torques because both have the same dependence on the angle of an in-plane magnetic field. Recent work by our group \cite{karimeddiny2020transverse} has shown that the established theory for the SP/ISHE voltage ($V_{\text{SP}}$) is effective at estimating the magnitude of this effect:
\begin{equation}
V_{\text{SP}}=\frac{e B_0 R_{\text{tot}}\theta_{SH}\cos^2(\phi)\sin(\phi)}{2\pi\alpha^2\gamma(2B_0+\mu_0 M_{\text{eff}})^2}g^{\uparrow\downarrow}_{\text{eff}}W\lambda_{\text{sd}}\text{tanh}\left(\frac{t_{\text{HM}}}{2\lambda_{\text{sd}}}\right)\left[(\tau^0_{\text{SH}})^2+\left(1+\frac{\mu_0 M_{\text{eff}}}{B_0}\right)(\tau^0_{\text{z}})^2\right]F_S(B)
\end{equation}
where $e$ is the charge of the electron, $B_0$ is the field at ferromagnetic resonance, $R_{\text{tot}}$ is the total resistance of the device, $\theta_{SH}$ is the ``internal" spin Hall ratio, $\phi$ is the in-plane angle of the magnetization with respect to the current flow axis, $\alpha$ is the Gilbert damping parameter, $\gamma$ is the electron gyromagnetic ratio, $\mu_0$ is the permeability of free space, $M_{\text{eff}}$ is the effective magnetization of the Fe layer, $g^{\uparrow\downarrow}_{\text{eff}}$ is the effective spin mixing conductance, $W$ is the width of the device, $\lambda_{\text{sd}}$ is the spin diffusion length of YbAl$_3$, $t_{\text{HM}}$ is the thickness of the YbAl$_3$ layer, $\tau^0_{\text{SH}}$ is the in-plane spin torque arising from the spin Hall effect, $\tau^0_{\text{z}}$ is the out-of-plane torque primarily arising from the Oersted field, and $F_S(B)$ is the symmetric Lorentzian function described in the main text. Some of the parameters in this equation can be determined directly from our ST-FMR measurements or other sample characterization. At 40 K and 23 GHz, for example, we have: $B_0$ = 0.17 T, $R_{\text{tot}}$= 83 $\Omega$, $\theta$ = $\frac{\pi}{4}$, $\alpha$ = 0.012, $\mu_0 M_{\text{eff}}$ = 2 T, $W$ = 10 $\mu$m, $t_{\text{HM}}$ = 6.7 nm, $\tau_{\text{SH}}$ = 0.02 GHz , and $\tau_{\text{z}}$= 0.1 GHz. However, the spin-mixing conductance, the spin diffusion length, and the spin Hall ratio are poorly-characterized even for well-studied materials, and are not known for YbAl$_{3}$/Fe.  

\begin{figure*}
\includegraphics[width=.8\textwidth]{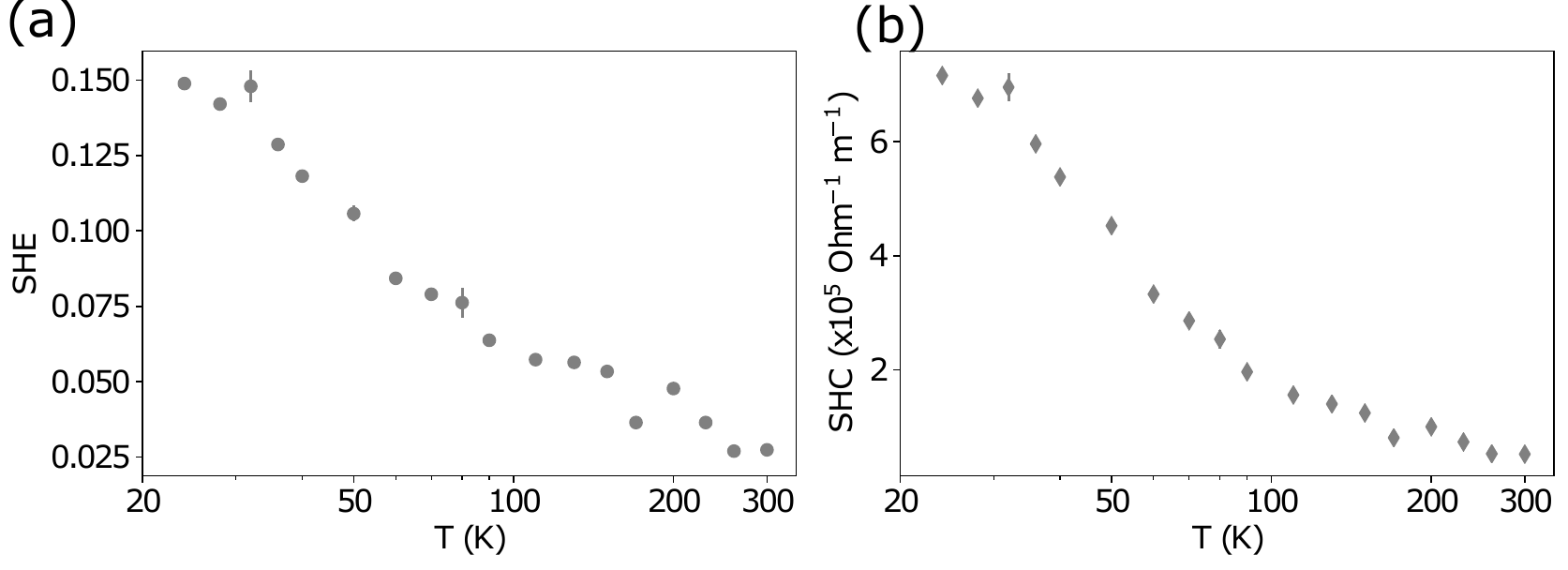}
\centering
   \caption{\textbf{Worst-case measured SHE and SHC due to spin-pumping.} Spin Hall efficiency (a) and spin Hall conductivity (b) of YbAl$_3$ sample assuming that the measured symmetric Lorentzian amplitude is due solely to SP/ISHE arising from the oscillation of the magnetization from the Oersted field torque, and that there is no interfacial spin memory loss.}
    \label{fig:ISHESP}
\end{figure*}

If we make the rough approximations that $g^{\uparrow\downarrow}_{\text{eff}} \approx$ 8 nm$^{-2}$, $\lambda_{\text{sd}} \approx 1$ nm (values comparable to Pt devices), and $\theta_{SH} \approx 0.4$, then Eq.~(S6) yields $V_{\text{SP}}\approx$ 0.5-1 $\mu$V. This is comparable to the magnitude of the signal we observe in our ST-FMR experiments (see Fig. 2(b) of the main text), and therefore likely not negligible.  It is therefore important to consider whether a contribution from this SP/ISHE signal could alter our conclusions about the strong temperature dependence of the spin torque conductivity from YbAl$_3$. We have checked this by considering a worst-case scenario, in which we assume that the entire symmetric resonance signal is due to the SP/ISHE effect arising from the magnetization being driven by the Oersted field torque. In this case, the magnitudes of the symmetric resonance signals correspond to a spin Hall conductivity that varies with temperature as shown in  Fig.~\ref{fig:ISHESP}. Therefore, the temperature dependence within this framework is even stronger than extracted from Fig. 2(c) of the main text, indicating that a possible contribution from the SP/ISHE does not invalidate our conclusion that the spin torque conductivity from YbAl$_3$ increases by at least a factor of 3 as a function of decreasing temperature from room temperature to the coherence temperature of YbAl$_3$.

\section{Magneto-Optical Kerr Effect (MOKE) Analysis}

As discussed in the main text, we use MOKE to measure the amplitude of current-induced magnetization oscillations as a function of position (scanning perpendicular to the current) for an 80 mT magnetic field applied both parallel and antiparallel to the current.  For the analysis, we separate the MOKE signals into the components that are even and odd with respect to the magnetic field.  We numerically integrate both signals as a function of position (after taking the absolute value of the even signal first) and extract the spin-orbit torque from the ratio of the integral of the odd signal to the integral of the even signal. Figure 3(b) shows the cumulative integral of each signal with the simulated expected lineshapes.

  The signal due to the in-plane spin-torques as a function of position across the device is given by the convolution of the beam shape with 
\begin{equation}
\begin{split}
V_d(B,x) = C\left\{ \begin{array}{lc}
      \xi_{\parallel,AD}+\xi_{\parallel,0}, \\
       		\text{for } B>0; -\frac{w}{2}<x<\frac{w}{2}\\
		\\
      -\xi_{\parallel,AD}+\xi_{\parallel,0},\\
        \text{for } B<0; -\frac{w}{2}<x<\frac{w}{2} \\
        \\
      0, \text{for } |x|>\frac{w}{2}
        \end{array}
        \right.
\end{split}
  \label{eq:mokerashba}
\end{equation}
where $V_d$ is the lock-in voltage, $B$ is the external magnetic field, $x$ is the position of the scan with $x$=0 corresponding to the middle of the device, and $C=K\frac{\hbar J_{c,YbAl_{3}}}{2eM_s t_{mag}(B+M_{\mathrm{eff}})}$. $K$ captures the Kerr constant and any other factors affecting the conversion of the optical response to a voltage signal from our split diode detector. The signal due to the Oersted field generated due to current flowing within the device is given by the convolution of the beam shape with
\begin{equation}
\begin{split}
V_s(B,x) = D\left\{\begin{array}{lc}
        \ln\left(\frac{t^2+(x-\frac{w}{2})^2}{t^2+(x+\frac{w}{2})^2}\right), & \text{for }|B| >0; -\frac{w}{2}<x<\frac{w}{2}\\
        0, & |x|>\frac{w}{2}
        \end{array}
        \right.
\end{split}
\label{eq:mokeoe}
\end{equation}
where $D=K\frac{\mu_0I_{total}}{2\pi w(B+M_{\mathrm{eff}})}$ and $\mu_0$ is the permeability of free space. Because we are measuring the out-of-plane component of the Oersted field, it is free from contamination due the non-Oersted out-of-plane spin-orbit torque that required careful consideration in the calibration of the ST-FMR signal. The signal due to $\xi_{\parallel,AD}$ is odd in external field while the signals due to the Oersted field and $\xi_{\parallel,0}$ are even. Thus we can isolate the individual signals by adding and subtracting scans taken at two equal and opposite values of the external applied field ($\pm$80 mT in our experiment). 

To extract values for the magnitude of the spin-orbit torques, we use that convolution is an area-preserving operation.  The integral $A_{d}$ as a function of position over the scan of Eq.~\ref{eq:mokerashba} is
\begin{equation}
    A_d \equiv C\xi_{\parallel,AD}w
\end{equation}
and the integral of the absolute value of $A_{s}$ over the scan of Eq.~(\ref{eq:mokeoe}) is
\begin{equation}
    A_s \equiv Dw\ln(4) + C\xi_{\parallel,0}w. 
\end{equation}
We take the absolute value before performing the spatial integral over $A_{s}$  because Eq.~(\ref{eq:mokeoe}) is odd about $x = 0$.  The integral of $|A_{s}|$ therefore gives a convenient overall measure of the scale of the out-of-plane magnetic field, from which to calibrate the charge current within the nonmagnetic layers. We subtract out any offset due to $\xi_{\parallel,0}$ from the even signal before integration to get just the integral of the Oersted signal $A_{Oe}\equiv D\times w\ln(4)$. Thus we can calculate
\begin{align}
\xi_{\parallel,AD}&=\frac{A_{d}}{A_{Oe}}\frac{\ln(4)D}{C}\\
&=\frac{A_{d}}{A_{Oe}}\frac{e}{\hbar}\frac{\mu_0 M_s t_{mag} t_{total}\ln(4)}{X\pi}
\end{align}
where $X$ is the fraction of the total current that flows through the YbAl$_3$ layer.

\section{Calculation of integrated spectral weight and background subtraction in YbAl$_{3}$ ARPES data}

Angle-resolved photoemission spectroscopy (ARPES) was performed using He I$\alpha$ photons from a VUV500 helium plasma discharge lamp in a measurement chamber with base pressure better than 5$\times$10$^{-11}$ Torr and equipped with a VG Scienta R4000 electron analyzer. Samples were transferred from the growth chamber into the measurement chamber under ultra-high vacuum conditions immediately after the completion of the growth process.

$E$-$k$ spectral maps are obtained along the $\Gamma$-X direction using He I$\alpha$ (21.2 eV) photon energy that corresponds to a $k_{z}$ value close to the bulk $\Gamma$ point \cite{chatterjee2017lifshitz}. Momentum integrated energy distribution curves (EDC) were obtained from the spectral maps by integrating over a momentum region $\pm (0.66 - 0.75) \pi/a$ as shown in Fig. 1(d) in the main text and also in Fig. S6(b). The momentum region was chosen to minimize contribution from non Yb 4$f$ derived bands, particularly the electron pocket near the $\Gamma$ point. We used a Shirley background \cite{shirley1972high} to estimate the contribution of inelastically scattered electrons.  Representative EDC plots before and after the background subtraction is shown in Figs. S6(c-e). Integrated spectral weight plotted in Fig. 1(j) of the main text is obtained by calculating the area under the curve over a binding energy region between -0.05 eV and 0.2 eV after background subtraction. Following the above procedure we obtain an enhancement ratio for the integrated spectral weight as $\frac{I(37 K)}{I(255 K)}$ = 2.2. Considering the intensity of the Kondo resonance (peak heights) in the EDC plots we obtain a ratio of 3.8. The obtained enhancement factor depends on the chosen binding energy range over which the spectral weight integration is performed and also on the details of the background subtraction and is found to vary between $\frac{I(37 K)}{I(255 K)}$ $\approx$ 2 and 4. Temperature dependent scaling behavior as described in the main text and saturation behavior at T$^{*}$ are found to be robust both to the choice of binding energy integration window and background subtraction procedure.

\newpage

\bibliography{SI.bib}
\newpage

\begin{figure}
\includegraphics[width=.7\textwidth]{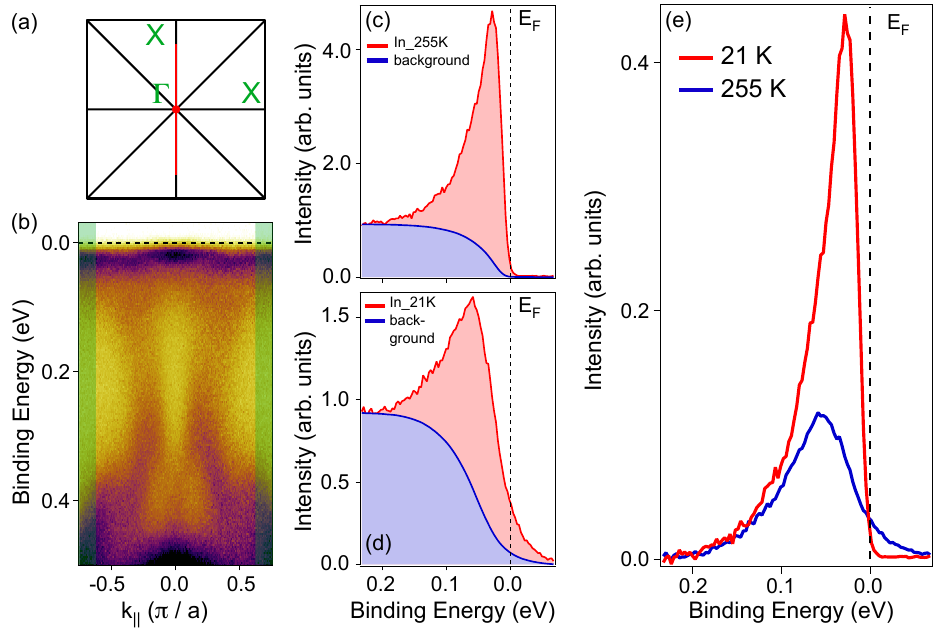}
\centering
   \caption{\textbf{Extraction of Yb 4$f$-derived band spectral weight from ARPES measurements.} (a) Surface Brillouin zone of YbAl$_{3}$ showing high symmetry points. (b) $E$-$k$ spectral map taken at 31 K corresponding to the ARPES cut (red line) shown in (a). Energy distribution curves taken at (c) 21 K and (d) 255 K integrated over the momentum region highlighted in green in (a). A Shirley background that takes into account contributions from inelastically scattered electrons is shown in blue. (e) Background-subtracted energy distribution curves at 21 K and 255 K.}
    \label{fig:SI5}
\end{figure}